# Increasing the Analytical Accessibility of Multishell and Diffusion Spectrum Imaging Data Using Generalized Q-Sampling Conversion

Fang-Cheng Yeh[1], and Timothy D. Verstynen[1]

[1]Department of Psychology and Center for the Neural Basis of Computation, Carnegie Mellon University, Pittsburgh, PA, USA;

*Correspondence to:

Timothy D. Verstynen Ph. D.,

Department of Psychology and Center for the Neural Basis of Computation,

Carnegie Mellon University, Pittsburgh, PA, USA

Email: timothyv@andrew.cmu.edu

ABSTRACT

Many diffusion MRI researchers, including the Human Connectome Project (HCP), acquire data using multishell (e.g., WU-Minn consortium) and diffusion spectrum imaging (DSI) schemes (e.g., USC-Harvard consortium). However, these data sets are not readily accessible to high angular resolution diffusion imaging (HARDI) analysis methods that are popular in connectomics analysis. Here we introduce a scheme conversion approach that transforms multishell and DSI data into their corresponding HARDI representations, thereby empowering HARDI-based analytical methods to make use of data acquired using non-HARDI approaches. This method was evaluated on both phantom and in-vivo human data sets by acquiring multishell, DSI, and HARDI data simultaneously, and comparing the converted HARDI, from non-HARDI methods, with the original HARDI data. Analysis on the phantom shows that the converted HARDI from DSI and multishell data strongly predicts the original HARDI (correlation coefficient > 0.9). Our in-vivo study shows that the converted HARDI can be reconstructed by constrained spherical deconvolution, and the fiber orientation distributions are consistent with those from the original HARDI. We further illustrate that our scheme conversion method can be applied to HCP data, and the converted HARDI do not appear to sacrifice angular resolution. Thus this novel approach can benefit all HARDI-based analysis approaches, allowing greater analytical accessibility to non-HARDI data, including data from the HCP.

# INTRODUCTION

Diffusion MRI offers a non-invasive way to map the structural connectivity of the human brain by tracking the local fiber orientations, reconstructed within each voxel, using a deterministic or probabilistic fiber tracking algorithm (Behrens et al., 2003; Behrens et al., 2007; Wedeen et al., 2012). While for most researchers interested in simple voxel-wise metrics of white matter integrity, diffusion tensor imaging remains the most common diffusion MRI approach, groups focused on improved fiber tracking capabilities will often adopt higher angular resolution approaches. For example, the Human Connectome Project (HCP) acquires diffusion MRI data using multishell (WU-Minn consortium; Sotiropoulos et al., 2013) and diffusion spectrum imaging (DSI) schemes (USC-Harvard consortium; McNab et al., 2013) approaches. One problem with these high resolution acquisition approaches is that they cannot be reconstructed by spherical deconvolution methods (Alexander, 2005; Dell'acqua et al., 2009; Tournier et al., 2008) and q-ball imaging (Descoteaux et al., 2007; Tuch, 2004). These two reconstruction methods are based on the high angular resolution diffusion imaging (HARDI) acquisition approach (Tuch et al., 2002) that collects a large number of diffusion images at different directions using a uniform strength of diffusion sensitization gradient (i.e., a single shell). These analytical limitations greatly hinder the application of multishell and DSI data to a wider user community that regularly uses HARDI–based approaches. While changes to HARDI algorithms may eventually allow for analysis of multishell data (Cheng et al., 2014; Jeurissen et al., 2014), these methods cannot be equally applied to DSI schemes. Moreover, these methods did not include the correction for the gradient nonlinearity (Sotiropoulos et al., 2013) that leads to variant gradient sensitization strength and direction. Therefore a new approach is needed to handle different diffusion sampling schemes and the gradient nonlinearity problem simultaneously.

Here we propose a scheme conversion method to transform non-HARDI data to a corresponding HARDI data set, thereby empowering all HARDI-based methods to make use of non-HARDI data. This method works by making use of the generalized q-sampling method (Yeh et al., 2010) that provides a linear relation between spin distribution functions (SDFs) and diffusion signals. This linear relation enables a scheme conversion while also incorporating the correction for gradient nonlinearity.

In order to evaluate the effectiveness of this scheme conversion approach, we conducted a phantom study and several in-vivo human studies. In the non-human phantom, HARDI, multishell, and DSI data were acquired and the multishell and DSI data were converted to a corresponding HARDI data set (hereafter referred to as the "converted HARDI" data set). A correlation analysis was conducted between the converted HARDI and the HARDI acquired from the MR scanner (termed "original HARDI" hereafter) to examine whether the converted HARDI correlates well with the original HARDI. In our first in-vivo study, we further applied constrained spherical deconvolution (CSD; Tournier et al., 2007) to the converted and original HARDI and examined whether the angular resolution of the converted HARDI can be compare to that of the original HARDI. We also calculated the connectivity matrices from converted and original HARDI data sets and determined their similarity using a correlation analysis. Finally, in a second in-vivo study we demonstrate the generalizability of our approach to data from both consortium sites of the Human Connectome Project (HCP). We examined whether the HCP data from WU-Minn and USC-Harvard consortiums can be converted to HARDI and reconstructed by CSD. The fiber orientations resolved by CSD were then compared with those from the ball-and-sticks model applied to the multishell data and those from the diffusion spectrum reconstruction. The utility of this conversion approach to HCP and non-HCP data will then be discussed.

**METHODS AND MATERIALS**

*Diffusion MRI scheme conversion*

Generalized q-sampling method provides a linear relation between diffusion MR signals and the spin distribution function (SDF; Yeh et al., 2010). This linear relation enables a direct conversion between SDFs and diffusion signals acquired from single-shell (HARDI), multishell, and grid (DSI) schemes. Since the SDFs from different schemes present a consistent pattern (Yeh and Tseng, 2013; Yeh et al., 2010, 2011), we can make use of the SDF to convert diffusion signals from one sampling scheme to another (Figure 1). DSI or multishell data can be converted to a common SDF and the linear relation between SDF and the HARDI signals will allow for estimating the corresponding HARDI representation by solving the inverse problem using constraint optimization.

To illustrate this idea, we start with the generalized q-sampling reconstruction, which is based on the linear relation between the diffusion MRI signals and the spin distribution function (SDF).

$$\pmb{\psi} = \mathbf{A} \cdot \mathbf{w} \qquad (1)$$

where $\pmb{\psi}$ is a diffusion ODF vector, and $\mathbf{w}$ is a vector of diffusion MRI signals, and its $i$-th dimension represents the diffusion signal acquired by a b-value of $b_i$ and diffusion gradient direction (b-vector) of $\hat{b}_i$. $\mathbf{A}$ is a matrix, and its element, $\mathbf{A}_{i,j}$, at row $i$ and column $j$ is defined as follows:

$$\mathbf{A}_{i,j} = \mathrm{sinc}\left(\sigma\sqrt{6D\cdot|\mathbf{b}_i|}\cdot\left\langle\frac{\mathbf{b}_i}{|\mathbf{b}_i|},\hat{\mathbf{u}}_j\right\rangle\right) \quad (2)$$

where $\sigma$ is a length ratio that controls the detection radius of the diffusion. $D$ is the diffusion coefficient of free water diffusion and $\hat{\mathbf{u}}_j$ is a unit vector representing the $j$-th direction of the diffusion ODF. Using Eq. (1), we can convert the DSI or multishell signals, $w$, to their corresponding HARDI representation, $w_h$, by equalizing their SDFs.

$$\mathbf{A}_h \cdot \mathbf{w}_h = \mathbf{A} \cdot \mathbf{w} \quad \text{s.t.} \quad \mathbf{w}_h \geq 0 \quad (3)$$

where $A_h$ is a matrix defined by a HARDI b-table, and $w_h$ is the corresponding HARDI representation to estimate. Eq. (3) formulates the conversion of the MRI signals as an inverse problem, and we can construct an over-determined equation (more equations than unknowns) by assigning more sampling directions in SDF than in HARDI. Eq. (3) can be solved by using the Tikhonov regularization.

$$\mathbf{w}_h \approx \left(\mathbf{A}_h^T\mathbf{A}_h + \lambda\mathbf{I}\right)^{-1}\mathbf{A}_h^T\mathbf{A}\cdot\mathbf{w} \quad (4)$$

where $\lambda$ is a regularization parameter and we chose the smallest possible $\lambda$ that resulted in more than 99% of the estimated HARDI representation being positive within the brain mask. This loosens the constraint for Eq. (3) and provides a quick estimation for the constraint optimization. The conversion routine was implemented in DSI Studio (http://dsi-studio.labsolver.org), a publicly available and open source tool.

*Correction for gradient nonlinearity*

The nonlinearity of diffusion gradients can induce a prominent image distortion that alters the effective b-values (Bammer et al., 2003; Sotiropoulos et al., 2013). This can be corrected using the nonlinear terms of the magnetic field obtained from gradient coils (Jovicich et al., 2006). The multishell data set from the WU-Minn consortium includes a 3-by-3 gradient deviation matrix **G** for each voxel to estimate the effective gradient direction and strength. This matrix can be directly embedded in the element of matrix A to account for the effect of gradient nonlinearity:

$$\mathbf{A}_{i,j} = \operatorname{sinc}\left( \sigma \sqrt{6D \cdot |\mathbf{Gb}_i|} \cdot \left\langle \frac{\mathbf{Gb}_i}{|\mathbf{Gb}_i|}, \hat{\mathbf{u}}_j \right\rangle \right) \quad (5)$$

Eq. (5) can replace Eq. (2), and the converted HARDI representation already considers the gradient nonlinearity.

*Diffusion phantom*

An anisotropic diffusion phantom (Brain Innovation, Maastricht, Netherlands; Pullens et al., 2010) was scanned using a 3T Siemens Verio scanner (Siemens, Erlangen, Germany) in the Scientific Imaging and Brain Research Center at Carnegie Mellon University. The phantom consists of one straight fiber bundle and a pair of crossing fiber bundles, as shown in Fig 2. Each bundle contains 10,000 polyester yarns (KUAG Diolen, 22 dtex 18) and each yarn is made up of 18 fibers. The production process and phantom property is detailed in the corresponding reference above. The acquisition parameters for HARDI, multishell, and DSI schemes are listed in Table 1. The images were acquired using the same spatial parameters: the field of view was 288 mm × 288 mm, the matrix size was 96 × 96, the slice thickness was 3.0 mm (no gap), resulting in a voxel size of 3.0 mm × 3.0 mm × 3.0mm. The multishell and DSI data were converted to 256-direction HARDI using a regularization parameter of 0.05. Both converted and original HARDI were multiplied by a constant that makes the mean of the overall signals equal to 0.5. To exclude background noise, we selected two regions of interest (ROIs), one placed on the straight fibers and another placed at the center of the crossing fibers. The diffusion signals at these two ROIs were averaged to increase the signal to noise ratio. Each diffusion weighted image had one average signal for the straight fibers, and one average signal for the crossing fibers. A total of 256 average values (one average value from each diffusion weighted image) were calculated for original HARDI. The same processing procedures were applied to the HARDI converted from DSI and multishell data. A correlation analysis between the converted and original HARDI was conducted to examine whether the converted HARDI can predict original HARDI.

*In vivo experiment*

We used publicly available data from Advanced Biomedical MRI Lab at National Taiwan University Hospital (http://dsi-studio.labsolver.org/download-images). The data include HARDI, multishell, and DSI data acquired on a 25-year-old male subject using a 3T MRI system (Tim Trio; Siemens, Erlangen, Germany). A 12-channel coil and a single-shot twice-refocused echo planar imaging (EPI) diffusion pulse sequence was used to acquire HARDI, multishell, and DSI data on the same subject, as summarized in Table 1. The HARDI, multishell, and DSI data were acquired using the same spatial parameters: the field of view was 240 mm × 240 mm, the matrix size was 96 × 96, the slice thickness was 2.5 mm (no gap), and the number of the slices was 40 to cover the cerebral cortex, resulting in a voxel size of 2.5 mm × 2.5 mm × 2.5 mm. The multishell and DSI data were converted to 252-direction HARDI using a regularization parameter of 0.05.

Both converted and original HARDI were multiplied by a constant that makes the mean of the overall signals equal to 0.5. The JHU white-matter tractography atlas (Laboratory of Brain Anatomical MRI, Johns Hopkins University) was used to segment corpus callosum, cerebral peduncle, coronal radiata, internal capsule, and cingulum pathways. Unlike what was done in the phantom study, we did not average the signals of each region because the containing fibers may not be oriented at the same direction. A correlation analysis between converted and original HARDI signals was conducted to examine whether the converted HARDI representation can predict those of the original HARDI.

We also calculated the fiber orientation distribution function (fODF) from the converted HARDI and examined whether they presented the same pattern as those from the original HARDI. The converted and original HARDI were reconstructed using constrained spherical harmonics (MRtrix, www.nitrc.org/projects/mrtrix) to obtain fODFs. As suggested in the MRtrix's user document, the response function was estimated using voxels with FA value greater than 0.7 and a maximum harmonic order of 8 was used. The calculated fODFs were compared to examine the consistency. A deterministic fiber tracking algorithm (Yeh et al., 2013) was applied to obtain a total of 1,000,000 tracks using an fODF threshold of 0.4. The connectivity matrices were calculated using the Automated Anatomical Labeling (AAL) atlas. A correlation analysis between matrices was conducted to examine whether the matrix from converted HARDI correlates well with the one from the original HARDI.

*Human Connectome Project Data*

A multishell data set was selected from the WU-Minn consortium (subject# 113619) and a DSI data set was from USC-Harvard consortium (subject# 20). The WU-Minn multishell data were acquired in a Siemens 3T Skyra scanner using a 2D spin-echo single-shot multiband EPI sequence with a multi-band factor of 3 and monopolar gradient pulse (Sotiropoulos et al., 2013). The spatial resolution is 1.25 mm isotropic. TR=5500 ms, TE=89ms. The b-values were 1000, 2000, and 3,000 s/mm$^2$. The total number of diffusion sampling directions was 270. The total scanning time was approximately 55 minutes. The multishell data were converted to HARDI using a regularization parameter of 0.05 and a b-value of 4,000 s/mm$^2$. The diffusion gradient nonlinearity was corrected using Eq. (5). The converted HARDI was analyzed using CSD implemented in MRtrix (http://www.nitrc.org/projects/mrtrix/). To examine the angular resolution of our converted HARDI, we compared the CSD results with the ball-and-sticks model (Behrens et al., 2003), the recommended analysis method for the WU-Minn HCP data (Sotiropoulos et al., 2013). The FSL's *bedpostx* program was used to resolve a maximum of 3 fibers resolved per voxel. All default parameters were used. The resolved fiber orientations were compared with those resolved by CSD applied to the converted HARDI.

The USC-Harvard DSI data were acquired in a Siemens 3T Connectome scanner equipped with a 300 mT/m gradient system (McNab et al., 2013). A 2D spin-echo EPI sequence was used to acquire diffusion images. The spatial resolution is 1.5 mm isotropic. TR=4200 ms, TE=53ms. The maximum b-value was 15,000 s/mm$^2$, and the total number of diffusion sampling directions was 515. The DSI data were converted to HARDI using a regularization parameter of 0.05 and a b-value of 4,000 s/mm$^2$. The converted HARDI was analyzed using CSD implemented in MRtrix (http://www.nitrc.org/projects/mrtrix/). The CSD results were compared with DSI reconstruction.

The converted HARDI of the recent HCP release (489 subjects from the WU-Minn consortium and 35 subjects from the USC-Harvard consortium) will be available at http://dsi-studio.labsolver.org/download-images

**RESULTS**

*Phantom study*

The converted HARDI data was found to be strongly correlated the original HARDI data in the phantom comparison study. Figure 3 shows the scatter plots of the raw diffusion MRI signals from the converted HARDI and original HARDI data sets. The regression equation and correlation coefficient are listed in Table 2. The high correlation coefficient (> 0.9) suggests that the converted HARDI is a good predictor of the original HARDI. It should be noted that the HARDI converted from the DSI data has a positive x-intercept. This may be due to the partial inclusion of the background free water diffusion, which gives substantial signal to the low b-value acquisition (< 1,000 s/mm$^2$) in DSI. The multishell and HARDI data do not include the low b-value acquisition and, thus, the diffusion signals are substantially attenuated.

*In-vivo experiment*

In the first human subject, the converted HARDI correlates well with the original HARDI. Figure 4 shows the HARDI images converted from multishell (Fig. 4A) and DSI (Fig. 4B) data, as well as the original HARDI image acquired from the MRI scanner (Fig. 4C). Since the multishell, DSI, and the original HARDI data were all acquired on the same subject we can directly compare voxelwise signal estimates. The images in Figure 4 present the axial views of the centrum semiovale. The converted HARDI shows a signal intensity pattern consistent with the original HARDI. The hyper-intensity regions are consistent in the converted and original HARDI. Of particular interest is the fact that the converted HARDI shows less noisy images compared to the original HARDI. This may be due to the regularization used in the scheme conversion that cancels some of the noise and improves the image quality.

As with the phantom study, in the human subject the converted HARDI data correlated highly with the original HARDI data at a voxelwise level. Figure 5 shows the scatter plots of the raw diffusion MRI signal from the HARDI converted from DSI and multishell data against those from the original HARDI. The regression equation and correlation coefficient are listed in Table 2. The correlation coefficients range from 0.67 to 0.84, suggesting a strong similarity in signal values. The HARDI converted from DSI shows unbiased regression equations with coefficients around 1.0 and intercepts around 0. However, the HARDI converted from multishell results in regression equations with coefficients around 0.8 and intercepts around 0.1. The bias may be due to the b-value difference between the multishell acquisition (b-values = 1500 and 3000 s/mm$^2$) and the original HARDI (b-value = 4000 s/mm$^2$). The multishell data were acquired by lower b-value and the signals were higher, resulting in biased coefficients.

We also found that the fODFs from the converted HARDI data are qualitatively similar to the fODFs from the original HARDI data. Figure 6 shows the results of CSD applied to the HARDI converted from the multishell scheme (Fig. 6A), the HARDI converted from the DSI scheme (Fig. 6B), and the original HARDI (Fig. 6C). The fODFs are presented in a coronal view focusing on the same slice covering the central semiovale. The fODFs of the converted and original HARDI present very similar shapes in these voxels. The crossing fibers formed by corpus callosum (horizontal) and corticospinal tracts (vertical) can be resolved using converted HARDI, though the converted HARDI was calculated using fewer diffusion sampling directions. This suggests that the converted HARDI can achieve an angular resolution comparable to the original HARDI.

The similarity in voxelwise fiber angle estimates suggests that these methods should also produce highly similar tractography results. Thus we compared the whole-brain connectivity results to a set white matter areas using deterministic tractograrphy (see Methods). The connectivity matrices calculated from converted and original HARDI were highly correlated. Figure 7 shows the connectivity matrix calculated from the HARDI converted from the multishell scheme (Fig. 7A), the HARDI converted from the DSI scheme (Fig. 7B), and original HARDI (Fig. 7C). These matrices present highly similar patterns of structural connectivity, with a correlation coefficient between the original HARDI and the HARDI converted from multishell data of r=0.8562 and a correlation coefficient between the original HARDI and the HARDI converted from DSI data of r=0.8884. This shows that the converted HARDI data can produce general structural connectivity estimates that are highly comparable to those from the original

HARDI.

*HCP data*

In a second in vivo study we illustrate that the scheme conversion can also be successfully applied to the DSI data set of the USC-Harvard consortium. Figure 8 shows the results of CSD applied to the converted HARDI. Fig. 8A shows a coronal view focused at the centrum semiovale, where corpus callosum (horizontal) crosses with corticospinal tracts (vertical) and arcuate fasciculus (through plane). The crossing of the fibers can be readily resolved using the fODFs calculated from CSD. Fig. 8B is an axial view focused at the centrum semiovale showing its crossing fibers. This result shows that CSD can be applied to DSI data from the HCP using the diffusion MRI scheme conversion. Similarly, we illustrate that the scheme conversion can be applied to the multishell data set of the WU-Minn consortium. Figure 9 shows the results of CSD applied to the HARDI converted from the multishell data set of the WU-Minn consortium. Fig. 9A and Fig. 9B are the coronal and axial views at the centrum semiovale, respectively.

We further used the multishell and DSI data of the HCP consortium to qualitatively examine the angular resolution. The converted HARDI were analyzed using CSD, and the resolved fiber orientations were compared with those from the ball-and-sticks model (multishell data) and DSI reconstruction (DSI data). Figure 10 compares the fiber orientations resolved by CSD applied to the HARDI converted from the multishell data (Fig. 10A) with the fiber orientations resolved by ball-and-sticks model applied to the same data (Fig. 10B). The figure is a coronal view focused on the centrum semiovale, whereas the horizontally going corpus callosum intersects with the vertically going corticospinal tracks and the superior longitudinal fasciculus that passes in-plane. The CSD and ball-and-sticks models resolve a similar pattern of crossing fibers, suggesting that the HARDI converted from the multishell scheme can achieve a comparable angular resolution. Likewise, Figure 11 compares the fiber orientations resolved by CSD applied to the HARDI converted from the DSI data (Fig. 11A) with the fiber orientations resolved by DSI reconstruction applied to the same data (Fig. 11B). The figure is also a coronal view focused on the centrum semiovale. The CSD and DSI resolve a similar pattern of crossing fibers, suggesting that the HARDI converted from the DSI scheme can achieve a comparable angular resolution.

**DISCUSSION**

Here we introduced a diffusion MRI conversion method that transforms multishell and DSI data sets into corresponding HARDI images in order to increase their analytical accessibility. Our phantom and in-vivo experiments show that the converted HARDI data are highly similar, in terms of voxelwise signal intensity, to the original HARDI images. We further applied CSD to the two converted HARDI data sets and found that the resulting fODFs and connectivity matrices were similar with those from original HARDI images. Moreover, we showed that our methods enable the use of CSD to reconstruct the multishell and DSI data from the HCP. Qualitative comparisons showed that CSD applied to the converted HCP data achieves an angular resolution comparable to ball-and-sticks model applied to the original HCP data.

The greatest advantage of this scheme conversion approach is that it bridges the gap between different diffusion sampling schemes that limits access to various popular reconstruction methods. By converting multishell and DSI data to HARDI, we can now make these data accessible to all HARDI-based methods, including q-ball imaging (Tuch, 2004) and deconvolution-based methods (Alexander, 2005; Dell'acqua et al., 2009; Tournier et al., 2007; Tournier et al., 2004). Non-HARDI analytical approaches (e.g., ball-and-stick) can still analyze HARDI data, and we only expand (not change) the scope of analysis that can be done here. Another advantage of this method is its ability to incorporate the correction of diffusion gradient nonlinearity into the scheme conversion calculation. Correcting for the remarkable changes in diffusion sensitization strength (b-value) and orientations caused by diffusion gradient nonlinearity has been a challenge for many diffusion MRI approaches. The elegance of our approach is that gradient nonlinearity correction is embedded in the conversion process of multishell and DSI data to a HARDI analog, and no additional routines are required. Thus, any HARDI reconstruction algorithm can be applied to converted HARDI data for which gradient nonlinearity artifacts have already been corrected.

While the improved analytical access of our approach has many benefits, we should point out that there are some limitations to our scheme conversion method. As shown in our phantom and in-vivo study, the diffusion images acquired from low b-value may have bias if they are converted to a high b-value scheme. This is due to the fact that low b-value acquisition is more sensitive to non-restricted diffusion, whereas high b-value acquisition is more sensitive to restricted diffusion. Although the conversion still resulted in a reasonably high correlation coefficient with the original HARDI data set, the SDFs from low b-value and high b-value may be different, and therefore converted HARDI images may have a signal bias. In addition, our approach is a one-way conversion, i.e., we cannot convert a HARDI data set into a multishell or DSI data set. Theoretically, a HARDI data set does not have the radial information that can differentiate slow diffusion from fast diffusion due to its uniform diffusion sensitization strength. Thus, the conversion from HARDI to multishell or DSI data set cannot fully recover this missing information. Finally, for the current study we only used a simple regularization approach to solve the constrained optimization problem in Eq. (3). This approach may not represent the most optimal solution to the parameters fit in Eq. (3). However, while there are more sophisticated solutions available to this optimization problem, they come at the expense of computation time. Future work should focus on finding the most optimal fitting approach that balances the accuracy and time trade-off.

Despite these current limitations we were able to show, for the first time, that a diffusion MRI scheme conversion method can convert multishell and DSI data, like those being collected as part of the HCP, to their corresponding HARDI images. The converted HARDI data can achieve an angular resolution comparable to original HARDI data set and the converted images can also be corrected for the diffusion gradient nonlinearity. This potentially benefits any HARDI-based method to make use of the HCP data, and similar non-HARDI acquisition projects, to reveal the complex axonal connections and to study the brain connectomics.

## ACKNOWLEDGMENTS

The authors thank Kevin Jarbo for thoughtful comments on the manuscript.

This research was sponsored by the Army Research Laboratory and was accomplished under Cooperative Agreement Number W911NF-10-2-0022. The views and conclusions contained in this document are those of the authors and should not be interpreted as representing the official policies, either expressed or implied, of the Army Research Laboratory or the U.S. Government. The U.S. Government is authorized to reproduce and distribute reprints for Government purposes notwithstanding any copyright notation herein. This research was supported by and NSF BIGDATA grant #1247658.

Part of the data were from the Advance biomedical lab at NTU, supported in part by the National Science Council, Taiwan (NSC100-3112-B-002-016, NSC100-2321-B-002-015), National Health Research Institute, Taiwan (NHRI- EX101-10145NI) and National Institutes of Health / National Institute of Mental Health, USA (U01MH093765).

Part of the data used this study were from (1) the Human Connectome Project, WU-Minn Consortium (Principal Investigators: David Van Essen and Kamil Ugurbil; 1U54MH091657) funded by the 16 NIH Institutes and Centers that support the NIH Blueprint for Neuroscience Research and by the McDonnell Center for Systems Neuroscience at Washington University and (2) the USC-Harvard Human Connectome Project (HCP) database (https://ida.loni.usc.edu/login.jsp). The HCP project (Principal Investigators : Bruce Rosen, M.D., Ph.D., Martinos Center at Massachusetts General Hospital; Arthur W. Toga, Ph.D., University of California, Los Angeles, Van J. Weeden, MD, Martinos Center at Massachusetts General

Hospital) is supported by the National Institute of Dental and Craniofacial Research (NIDCR), the National Institute of Mental Health (NIMH) and the National Institute of Neurological Disorders and Stroke (NINDS).

**Table 1. Summary of diffusion scans in phantom and in-vivo study**

|  | scheme | b-value (s/mm$^2$) | number of diffusion gradient encodings | TR | TE |
|---|---|---|---|---|---|
| Phantom study | | | | | |
| | two-shell | 1500/3000 | 30/64 | 2100/2500 | 135/143 |
| | grid (DSI) | 400~4000 | 515 | 1800 | 135 |
| | single-shell (HARDI) | 3000 | 256 | 2500 | 143 |
| In-vivo study | | | | | |
| | two-shell | 1500/3000 | 30/64 | 5500/6300 | 101/121 |
| | grid (DSI) | 400~4000 | 202 | 7200 | 144 |
| | single-shell (HARDI) | 4000 | 252 | 7200 | 133 |

Table 2. Results of correlation analysis in the phantom and in-vivo study

| | Original Data | Location | Regression Line | Correlation Coefficient |
|---|---|---|---|---|
| Phantom study | | | | |
| | DSI | Straight fibers | y = 1.7621 x - 0.3810 | 0.9576 |
| | DSI | Crossing fibers | y = 1.5545 x - 0.2772 | 0.9766 |
| | Multishell | Straight fibers | y = 1.0467 x -0.0234 | 0.9773 |
| | Multishell | Crossing fibers | y = 0.8970 x + 0.0515 | 0.9866 |
| In-vivo study | | | | |
| | DSI | corpus callosum | y = 1.1045 x - 0.0435 | 0.8251 |
| | DSI | cerebral peduncle | y = 0.9628 x + 0.0303 | 0.7912 |
| | DSI | coronal radiata | y = 0.9595 x + 0.0225 | 0.6777 |
| | DSI | internal capsule | y = 0.9756 x + 0.0258 | 0.7338 |
| | DSI | cingulum | y = 0.9067 x + 0.0099 | 0.6408 |
| | Multishell | corpus callosum | y = 0.8335 x + 0.0795 | 0.8476 |
| | Multishell | cerebral peduncle | y = 0.7951x + 0.1128 | 0.7996 |
| | Multishell | coronal radiata | y = 0.7812 x + 0.1148 | 0.7207 |
| | Multishell | internal capsule | y = 0.8073 x + 0.1228 | 0.7617 |

| | | | |
|---|---|---|---|
| Multishell | cingulum | y = 0.7317 x + 0.0925 | 0.6735 |

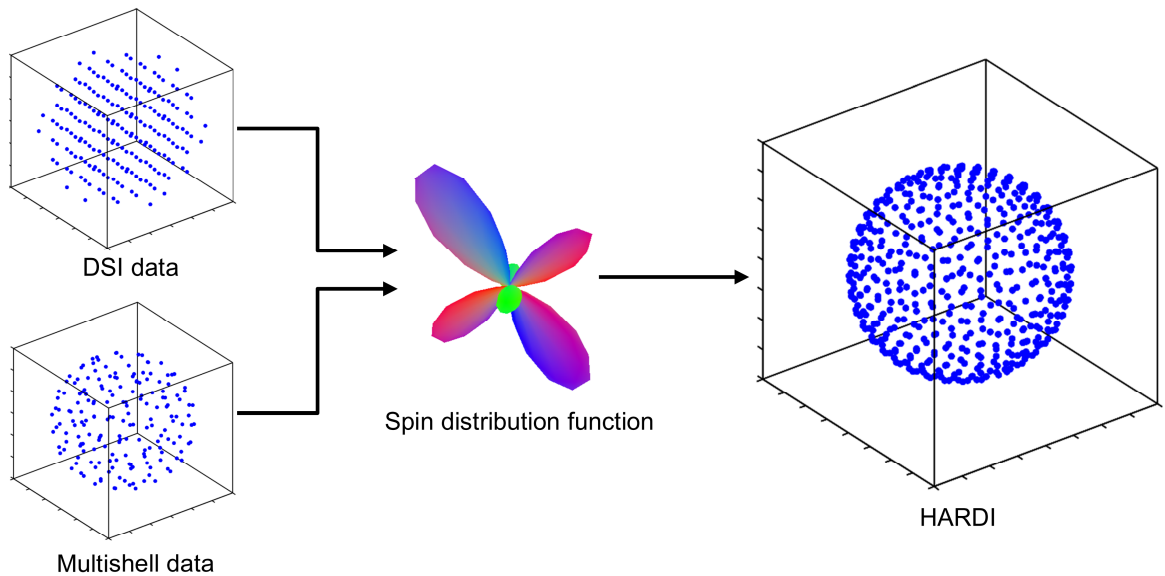

Figure 1. The scheme conversion method uses the spin distribution function (SDF) to convert multishell or DSI data to their corresponding HARDI representation. This is made possible by the linear relationship between the diffusion signals and the SDF provided by the generalized q-sampling reconstruction approach.

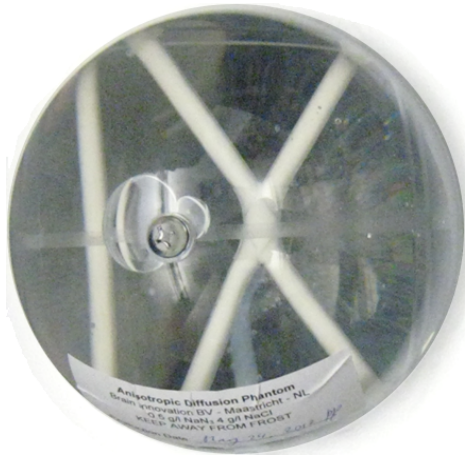

Figure 2. The layout of the anisotropic diffusion phantom. The phantom consists of one straight fiber bundle and a pair of crossing bundles.

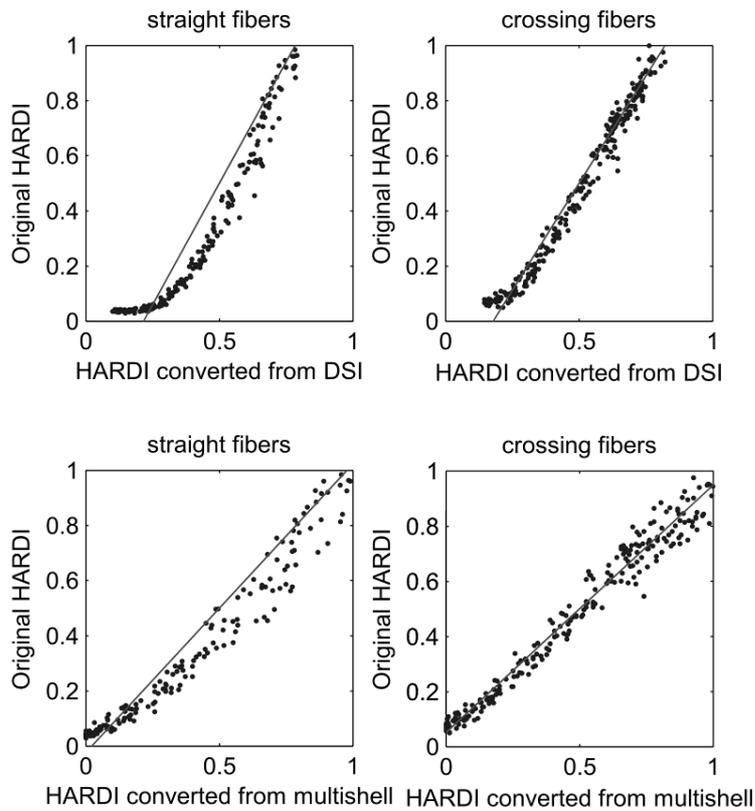

Figure 3. Scatter plots showing signal correlation between the converted HARDI and the original HARDI in our phantom study. The converted HARDI was calculated from the multishell or DSI data, whereas the original HARDI was acquired from the MR scanner. All comparisons had a high correlation coefficient (r > 0.9), suggesting that the converted HARDI data sets can be used as a proxy for the original HARDI images.

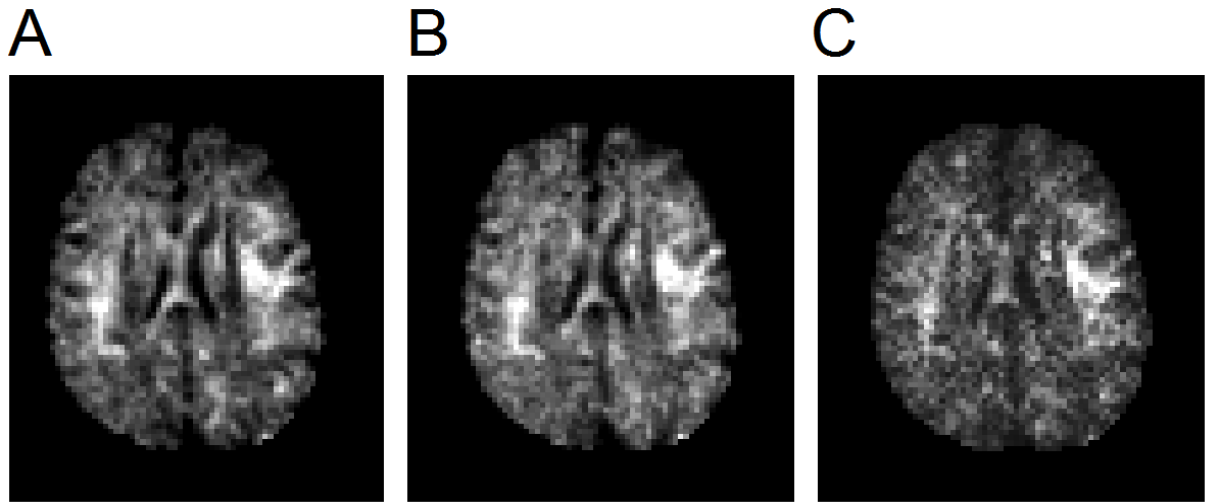

Figure 4. The HARDI images converted from (A) a multishell scheme and (B) a DSI scheme compared with (C) the original HARDI images from scanner. The converted HARDI images show a signal pattern similar to the original HARDI acquired from the MRI scanner.

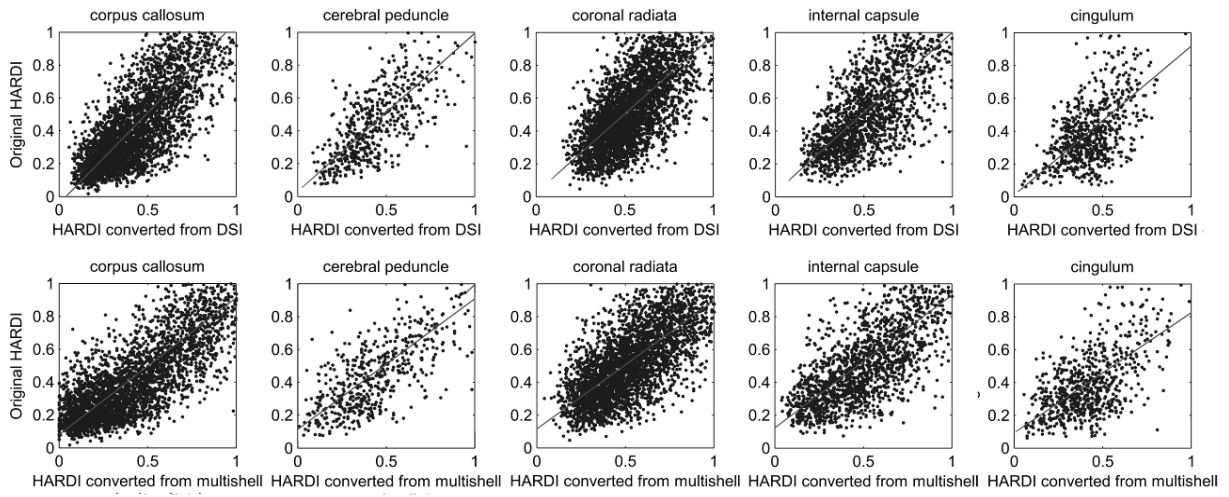

Figure 5. Scatter plots showing signal correlation between the converted HARDI and the original HARDI in our in-vivo study. The JHU white matter atlas was used to selected regions of interests in the analysis. The converted HARDI data correlates well with the original HARDI images (r > 0.6).

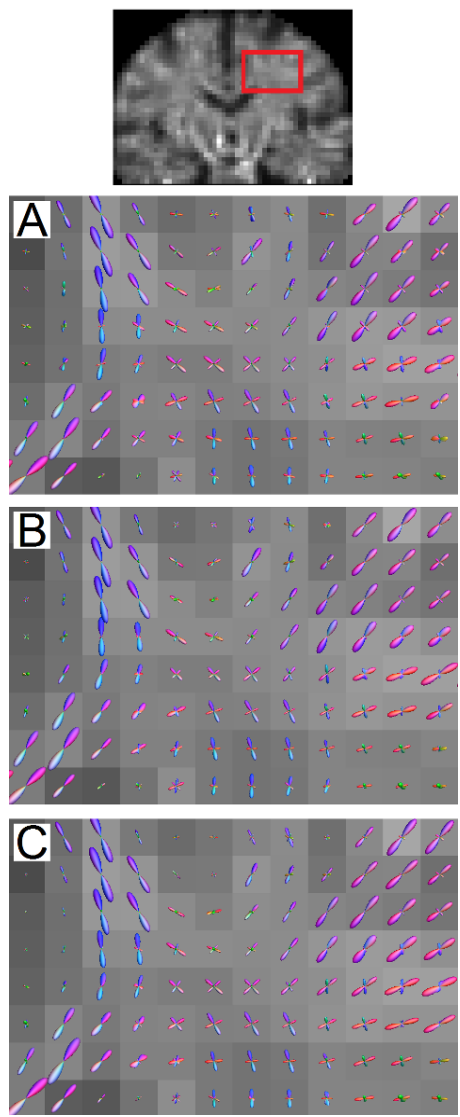

Figure 6. The results of CSD applied to (A) the HARDI converted from multishell data set, (B) the HARDI converted from DSI data set, and (C) the original HARDI data set. The fiber orientation distribution functions (fODFs) show similar patterns and comparable angular resolution.

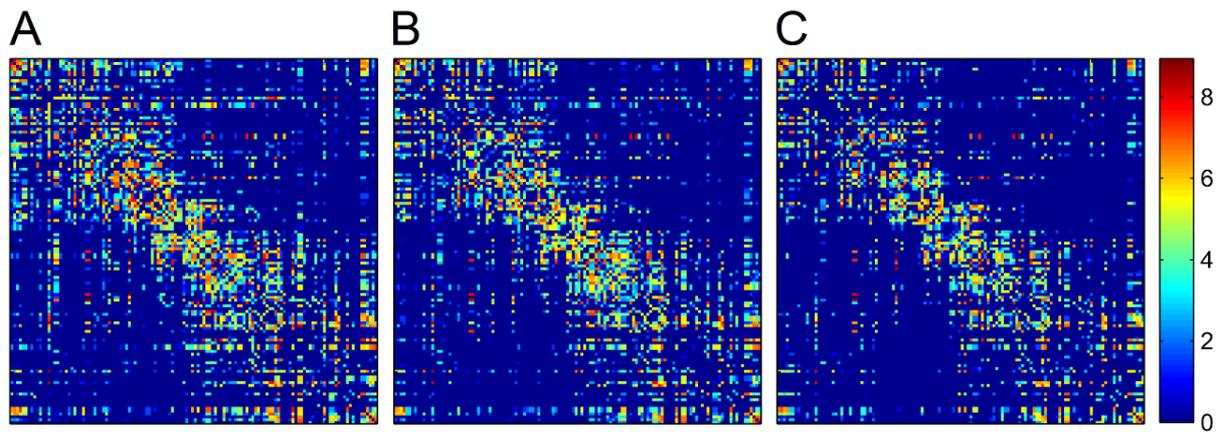

Figure 7. The connectivity matrices calculated from (A) the HARDI converted from multishell data set, (B) the HARDI converted from DSI data set, and (C) the original HARDI data set. These matrices show high correlation coefficient (all r's > 0.8).

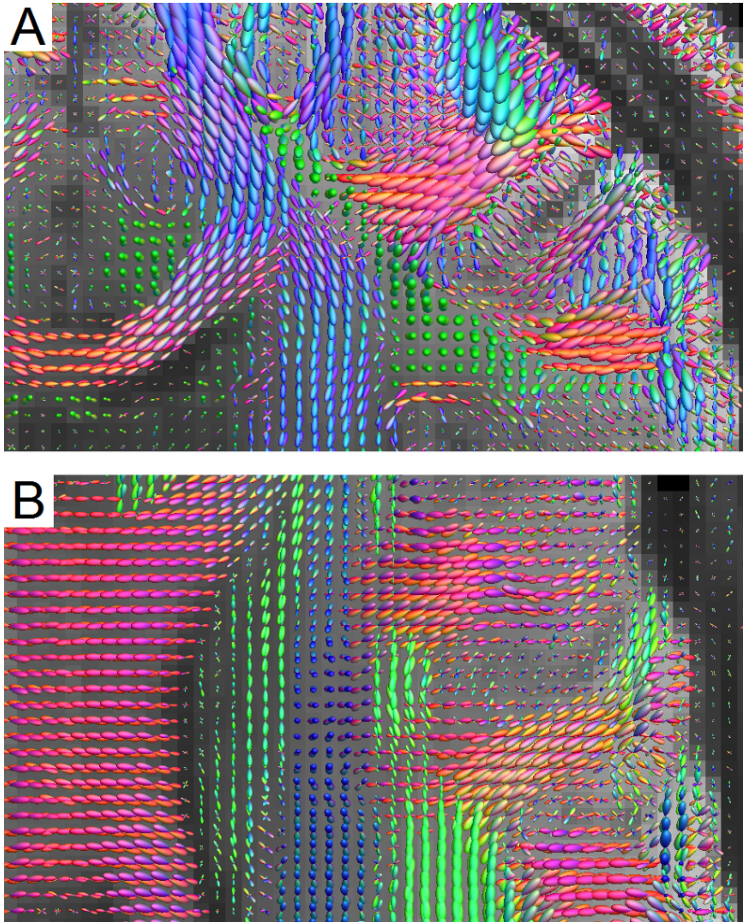

Figure 8. The results of CSD applied to the DSI data set of the USC-Harvard consortium using our scheme conversion method. The fODFs are presented by (A) a coronal view and (B) an axial view at the centrum semiovale.

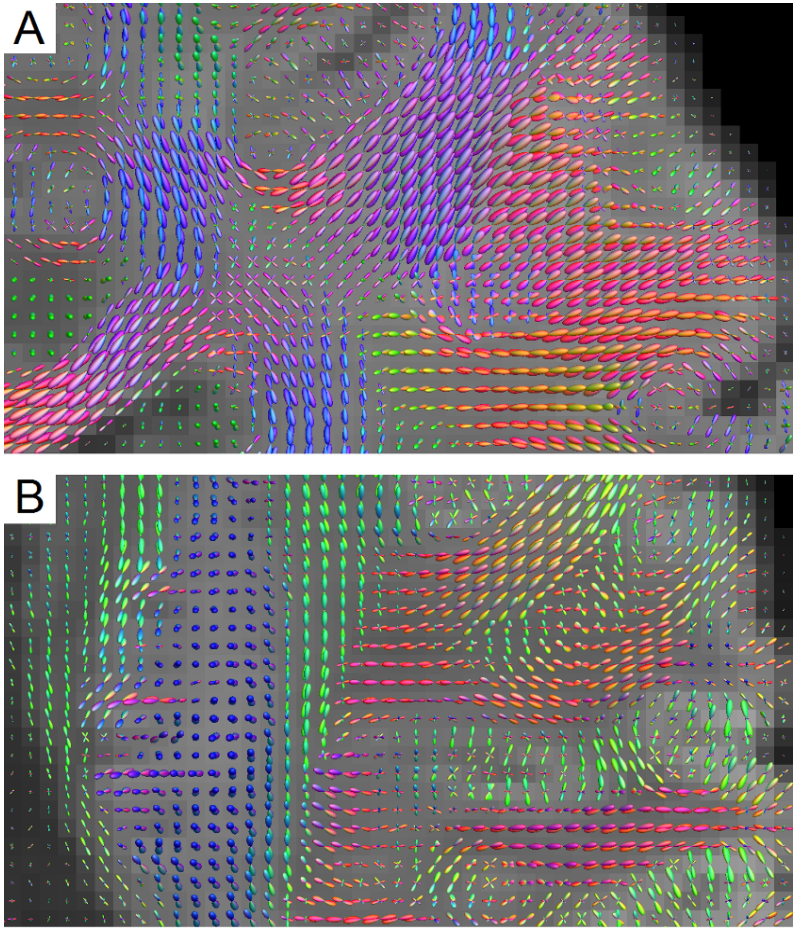

Figure 9. The results of CSD applied to the multishell data set of the WU-Minn consortium using our scheme conversion method. The fODFs are presented by (A) a coronal view and (B) an axial view at the centrum semiovale.

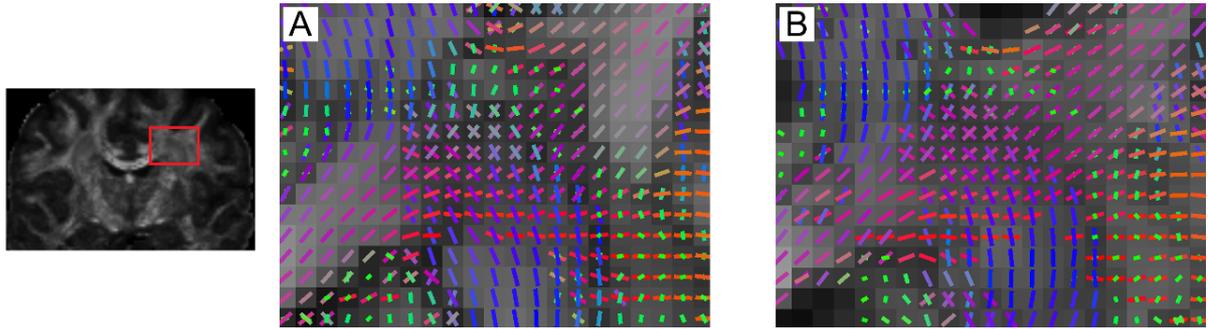

Figure 10. The fiber orientations resolved by (A) CSD applied to the multishell data of the WU-Min consortium using our scheme conversion method and (B) ball-and-sticks model applied to the same data. They achieve a comparable angular resolution, and the crossing fibers can be readily resolved.

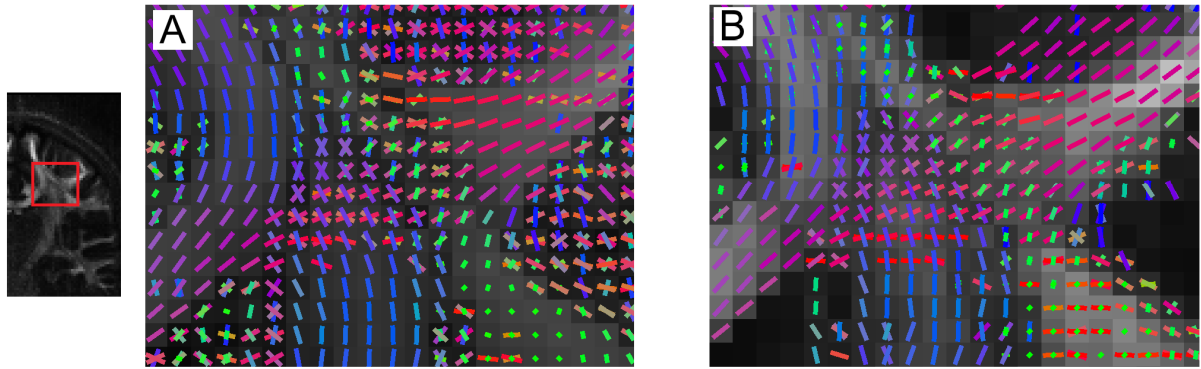

Figure 11. The fiber orientations resolved by (A) CSD applied to the DSI data of the USC-Harvard consortium using our scheme conversion method and (B) DSI reconstruction applied to the same data. Similar to the multishell data, here the data converted from DSI achieve a comparable angular resolution, and the crossing fibers can be readily resolved.